# Inferring entropy production rate from partially observed Langevin dynamics under coarse-graining

Aishani Ghosal[a] and Gili Bisker *[a,b,c,d]

The entropy production rate (EPR) measures time-irreversibility in systems operating far from equilibrium. The challenge in estimating the EPR for a continuous variable system is the finite spatiotemporal resolution, and the limited accessibility to all of the nonequilibrium degrees of freedom. Here, we estimate the irreversibility in partially observed systems following oscillatory dynamics governed by coupled overdamped Langevin equations. We coarse-grain an observed variable of a nonequilibrium driven system into a few discrete states, and estimate a lower bound on the total EPR. As a model system, we use hair-cell bundle oscillations, driven by molecular motors, such that the bundle tip position is observed but the positions of the motors are hidden. In the observed variable space, the underlying driven process exhibits second-order semi-Markov statistics. The waiting time distributions (WTD), associated with transitions among the coarse-grained states, are non-exponential and convey the information on the broken time-reversal symmetry. By invoking the underlying time-irreversibility, we calculate a lower bound on the total EPR from the Kullback-Leibler divergence (KLD) between WTD. We show that the mean dwell-time asymmetry factor – the ratio between the mean dwell-times along the forward direction and the backward direction, can qualitatively measure the degree of broken time reversal symmetry and increases with finer spatial resolution. Finally, we apply our methodology to a continuous-time discrete Markov chain model, coarse-grained into a linear system exhibiting second-order semi-Markovian statistics, and demonstrate the estimation of a lower bound on the total EPR from irreversibility manifested only in the WTD.

## 1 Introduction

Irreversible processes in living systems lead to the production of entropy, which is a measure of energy dissipation and a signature of the arrow of time.[1–5] Quantifying the entropy production can shed light on the underlying nonequilibrium dynamics and provide insights on the thermodynamic burden of biological processes.[6–8] There are primarily two methods to infer that a system is out-of-equilibrium: (i) invasive methods,[9–12] and (ii) non-invasive[13–16] methods. In invasive methods, the system's response to a perturbation is measured following an external manipulation, and the violation of the fluctuation-dissipation theorem (FDT)[9,17–22] confirms the nonequilibrium nature of the underlying process. On the other hand, non-invasive methods do not require a direct perturbation to a system, and can detect the nonequilibrium nature of the process from various system properties, such as broken time-reversal symmetry,[23,24] presence of net probability current of observables,[7,13,16,25–29] or asymmetric probability density function (PDF) of the timing of maximal observable values.[30]

One can estimate the EPR for discrete[31] and continuous systems[32–34] given that all out-of-equilibrium system variables are accessible; otherwise, the EPR estimation becomes challenging,[35–39] and the best estimate would be a lower bound on the total EPR value. Several studies focused on the fluctuations of the EPR calculated from partial information.[40–48] The mathematical relations that bound the EPR using the fluctuations of time asymmetric and generic variables are known as the thermodynamic uncertainty relation (TUR)[49–54] and kinetic uncertainty relation (KUR)[55], respectively. These relations have also been generalized for semi-Markov processes.[56,57] Recently, a unified relation considering both thermodynamic and kinetic quantities has been proposed.[58] For systems with partial information, estimators like the passive partial entropy production [47,59,60] and the informed partial entropy production[59–62] are helpful to get a dissipation bound; however, these fail to provide a tight bound on the total EPR for vanishing net current. These average partial entropy production estimators satisfy fluctuations theorems, and as such, they can be derived as a Kullback-Leibler divergence between the forward trajectory and the backward trajectory under auxiliary dynamics.[59]

The k-variable irreversibility measure is defined as, $\sigma_k \equiv k_B \lim_{t \to \infty} \frac{1}{t} D[P(\Gamma^k)||\tilde{\Gamma}^k)]$, where $k_B$ is the Boltzmann constant, $D[p||q]$ denotes the Kullback-Leibler divergence (KLD)[63,64] between two probability distributions $p$ and $q$, defined by $D[p||q] = \int dx\, p(x) \log(p(x)/q(x))$ and calculated on the positive support. [64] It is a measure of distinguishability[65] between two probability distributions, being non-negative in general, and zero for identical distributions. $\Gamma^k$ denotes the forward path of $k$ nonequilibrium variables for a time duration $t$, whereas $\tilde{\Gamma}^k$ denotes the corresponding backward path. Owing to the chain-rule of the relative entropy,[66] the more nonequilibrium variables (larger $k$) included in the path probability measure, the better the KLD bound is, i.e., $0 \le \sigma_1 \le \cdots \le \sigma_k \le \sigma_{k+1} \le \cdots \le \sigma_{tot}$, where $\sigma_{tot}$ is the total EPR calculated by the KLD between the forward and reverse trajectories with all the nonequilibrium degrees of freedom.[67,68] Obtaining a tight bound for a continuous variable system using the KLD estimator is challenging since some of the nonequilibrium variables may be inaccessible and sampling the distribution of paths becomes difficult.

In a recent study,[69] Roldan et al. transformed the forward and backward time series data of an observed variable of a continuous hair bundle system into two independent and identically

a. Department of Biomedical Engineering, Tel Aviv University, Tel Aviv 6997801, Israel
b. Center for Physics and Chemistry of Living Systems, Tel-Aviv University, Tel Aviv 6997801, Israel
c. Center for Nanoscience and Nanotechnology, Tel-Aviv University, Tel Aviv 6997801, Israel
d. Center for Light-Matter Interaction, Tel-Aviv University, Tel Aviv 6997801, Israel

distributed time series using a whitening approximation to estimate the KLD from two univariate distributions. They first calculated the EPR bound using only the observed degree of freedom, *i.e.*, the tip position of the hair bundle. Moreover, they used the finite time thermodynamic uncertainty relation to obtain a lower bound on the total EPR using two observables, the tip position and the transduction current, and found a better lower bound on the EPR compared to the one calculated using only one variable, as expected. The EPR estimate calculated with only one observed degree of freedom was typically three orders of magnitude smaller than the total EPR. However, using two observables and the TUR their measure was three orders of magnitude better than their single-variable result for the oscillatory regime and few fold smaller than the total EPR, but in the quiescent regime the result was three orders of magnitude smaller than the total EPR.

An estimator based on the KLD between waiting time distributions of the time forward and the time backward transitions between discrete states was shown to provide a lower bound on the total EPR,[62] given that the time-reversal operator does not lead kinetic hysteresis.[70–72] Applied to a semi-Markov process, or a continuous time random walk (CTRW), this KLD estimator of the EPR breaks into two contributions,[62] the affinity EPR, $EPR_{aff}$, which accounts for the net flux and affinity or the thermodynamic force,[66,68] and the waiting-time-distribution (WTD) EPR, $EPR_{WTD}$, which accounts for the broken time-reversal symmetry in the waiting time distributions.[62] In a semi-Markov process, which can originate from a coarse-grained Markovian system, the waiting time distribution does not follow an exponential distribution. If the coarse-graining procedure commutes with the time-reversal, the $EPR_{WTD}$ vanishes for semi-Markov processes.[70–73] However, for second-order semi-Markov processes, which naturally emerge when "lumping" adjacent states,[62,74] the $EPR_{WTD}$ can provide a lower bound on the total EPR, even when the system does not have any net current observed and $EPR_{aff} = 0$. A second-order semi-Markov process is considered a process where the waiting time distribution from a state $j$ to the state $k$ depends on the state $i$ visited prior to the state $j$. The term "second" comes from the fact that one has to consider doublets of states$[i, j]$ instead of just the current state $j$ to understand the future system dynamics. Describing processes by transitions instead of states,[75] the KLD estimator for the EPR was further applied to waiting times in between observed transitions.[73,76]

Skinner *et al.* presented new estimators to obtain the lower bound on the entropy production rates using optimization techniques.[77,78] They found an estimator given observables characterizing one-step transitions and two successive transitions, whereas in another publication the authors proposed an estimator given the observed waiting time distributions.[78]

There are several studies on the effect of coarse-graining (CG) on the EPR estimation[79,80,89–92,81–88] *specifically* discussing whether the CG procedure preserves the EPR fluctuations or not. EPR fluctuations can be preserved in the presence of time scale separation.[92] Given a proper decimation process, the coarse-graining of a Markov jump process can preserve the EPR fluctuations if the contribution of an erased loop to the entropy production is accounted for by self-jumps in the coarse-grained trajectory.[81] In contrast, many coarse-graining procedures do not preserve the EPR fluctuation, including (i) at the overdamped limit,[82,88] (it was proved earlier that the overdamped description provides the same EPR bound as that of the underdamped description only at high viscosity coefficient limit [83]), (ii) temporal coarse-graining of the system into $n$ equal time internals, in which the EPR estimation varies as $n^{-2}$,[84] (iii) considering the effect of fast or hidden degrees of freedom as a memory term,[85] (iv) coarse-graining the fast degrees of freedom resulting in an effective Markov process,[86,89–91] (v) altering the network topology or losing a fundamental cycle,[87] etc. In a recent study, using a Markovian model of a driven molecular motor, Hartich *et al.* compared different coarse-graining schemes, "milestoning" and "lumping", and found that the "milestoning" method can restore Markovian dynamics in the case of time-scale separation and preserves local detailed balance.[74,92]

The quantitative effect of the coarse-graining on the EPR was estimated in an experimental system of steady-state trajectories of a microtubule length using an optimization procedure of a two-step estimator, where it was demonstrated that increasing the spatial and temporal resolution while coarse-graining leads to an improved EPR bound.[77] Moreover, a recent study by Tan *et al.* [93] has found that the time-irreversibility varies non-monotonically with the lag time, *i.e.*, the time intervals between the position measurements, which determines the dissipation timescale.[93]

Here, we quantify the irreversibility using a non-invasive method to provide a lower bound on the total EPR in a partially observed model system with continuous variables following oscillatory dynamics, where one of its observables is coarse-grained into a few discrete states. We simulate an oscillating hair-bundle model in which the bundle's tip position is experimentally observed, whereas the position of the molecular motor is hidden. The coarse-grained process follows second-order *semi-Markov* statistics in the *reduced state space* (tip position variable space). In this model, the affinity entropy production contribution vanishes; therefore, the irreversibility information can only be accessed from the asymmetries of the waiting time distributions of the forward and the reversed transitions. After the decimation, we exploit the underlying broken time-reversal symmetry stemming from the difference in the PDFs of the waiting times for the upward and the corresponding downward transitions among different coarse-grained states, to calculate the EPR bound, $EPR_{WTD}$, by applying the KLD estimator. We show that the ratio of the means of the dwell time PDFs of the forward and reverse trajectories, termed the mean dwell-time asymmetry factor, can qualitatively detect the broken time reversal symmetry, and its variation with the number of coarse-grained states is studied. We further calculate the ratio between the $EPR_{WTD}$ and the total EPR as a function of the number of coarse-grained states to evaluate the tightness of the lower bound, and find that with finer resolution (larger number of coarse-grained states), the $EPR_{WTD}$ provides a better lower bound on the dissipation rate.

The paper is organized as follows. First, we introduce the model system and outline the calculation of the total EPR. Then, we describe our coarse-graining procedure, second-order semi-Markovian dynamics of the coarse-grained system, different contributions to the EPR, and mean dwell-time asymmetry factor in the next section. Subsequently, the effect of coarse-graining on the broken time-reversal symmetry, the EPR estimate, and the tightness of the lower bound as a function of number of coarse-grained states are discussed. Finally, we summarize and provide a future outlook.

## 2 Model System

We estimate the entropy production rate in a partially observed system described by a Langevin equation. To do so, we consider a model which captures the experimental observation of spontaneous oscillations of mechanosensory hair bundles of auditory hair cells.[69,94–98] These oscillations help to amplify the sound stimuli in the ear of vertebrates, and provide sensitivity and frequency selectivity. Moreover, these oscillations are known as

"active" oscillations, and they are distinct from "passive" oscillations that are obtained by blocking the corresponding transduction ion channels.[69] The activity originates from various molecular motors, which cannot be experimentally accessed. However, another degree of freedom coupled to the activity of the molecular motors – the tip position of the hair bundle ($X_1$) is experimentally observed. Due to the presence of activity, the system is out-of-equilibrium, and its dynamics is governed both by a conservative force $V(X_1, X_2)$, where $X_2$ represents the position of the center of mass of the molecular motors, and a non-conservative driving force, $F_{act}(X_1, X_2)$. The system can be described by the following coupled stochastic differential equations [69,94–96]

$$\lambda_1 \dot{X}_1 = -\frac{\partial V(X_1,X_2)}{\partial X_1} + \sqrt{2k_B T \lambda_1}\xi_1 \quad (1)$$

$$\lambda_2 \dot{X}_2 = -\frac{\partial V(X_1,X_2)}{\partial X_2} - F_{act}(X_1, X_2) + \sqrt{2k_B T_{eff} \lambda_2}\xi_2 \quad (2)$$

where $\lambda_1$ and $\lambda_2$ are the friction coefficients of the hair bundle tip and the molecular motor, respectively, $T$ and $T_{eff}$ are the environment temperature and the effective temperature characterizing the motor fluctuations, respectively, with ratio $T_{eff}/T > 1$. $\xi_1$ and $\xi_2$ are two independent white noise terms with zero-mean and correlation $\langle \xi_i(t)\xi_j(t') \rangle = \delta_{ij}\delta(t-t')$, and $k_B$ is the Boltzmann constant. The functional form of the conservative force, $V(X_1, X_2)$, which is proportional to the difference between the positions of the coupled variables[69,94–96], is:

$$V(X_1, X_2) = \frac{k_{gs}\Delta X^2 + k_{sp}X_1^2}{2} - Nk_B T \ln\left[e^{\left(\frac{k_{gs}D\Delta X}{Nk_B T}\right)} + A\right] \quad (3)$$

where $k_{gs}$ and $k_{sp}$ are the stiffness coefficients, $\Delta X = X_1 - X_2$ is the separation between the position of the hair bundle and the molecular motors, $D$ is the gating swing, and $N$ is the number of transduction channels. $A = \exp[(\Delta G + (k_{gs}D^2)/2N)/(k_B T)]$, and $\Delta G$ is the energy difference between the open and closed states of the ion channel. The active non-conservative force exerted by the molecular motors is defined by $F_{act}(X_1, X_2) = F_{max}(1 - SP_0(X_1, X_2))$. The probability of the transduction channel being open is $P_0(X_1, X_2)$, and is defined by $P_0(X_1, X_2) = 1/[1 + A\exp(-k_{gs}D\Delta X/Nk_B T)]$. The non-conservative force depends on the maximum motor force acting on the system ($F_{max}$), and the calcium-mediated feedback strength ($S$). The main sources of the non-equilibrium drive come from the ratio $T_{eff}/T$ being greater than unity, and the maximal force ($F_{max}$) exerted by the molecular motors. This model[69,94–96] was shown to agree well with experimental results.

First, we numerically solve the coupled differential equations (Eq. 1 and Eq. 2) for a fixed ratio between the effective temperature and the temperature of the environment ($T_{eff}/T = 1.5$), and different values of $S$ (0.5, 1, 1.5) and $F_{max}$ (70 pN, 80 pN, 90 pN) to obtain simulated trajectories of the hair bundle tip position and the motor position (see Figure 1 for details on all the parameters used). Although there is clearly a directional current in the $X_1$ - $X_2$ plane (Figure 1a) manifesting the nonequilibrium nature of the process, its signature is not obviously present in the trajectories of $X_1$ or $X_2$ as a function of time, which oscillate around their respective mean values (as shown in Figure 1b and Figure 1c) for a particular set of the driving parameter values, and Supplementary Information, Figure S1 for additional realizations with different parameters).

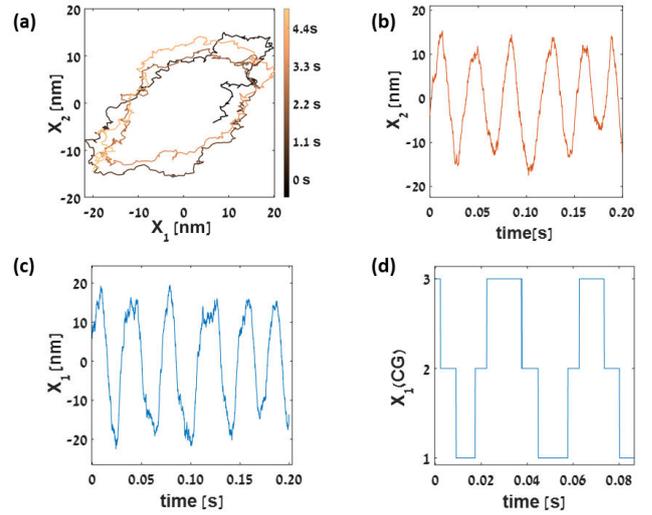

Figure 1 Simulated trajectories of the observed ($X_1$, the tip position of the hair bundle) and the hidden ($X_2$, the position of the molecular motors) variables and the coarse-grained trajectory of the observed variable after spatial coarse-graining (a) The trajectories in the $X_1 - X_2$ plane for fixed values of driving parameters: $F_{max} = 70\ pN$, $S = 1$, $T_{eff}/T = 1.5$). The color of the curve represents time going from dark to bright (b) $X_2 = X_2 - \langle X_2 \rangle$, as a function of time for fixed values of driving ($F_{max} = 70\ pN$, $S = 1$, $T_{eff}/T = 1.5$). (c) $X_1 = X_1 - \langle X_1 \rangle$, as a function of time for the same values of driving parameters, which does not show any sign of net flux, (d) The coarse-grained trajectory for 3 CG states at above-mentioned parameter values. All the quantities plotted are calculated for the following additional parameter values: $\lambda_1 = 2.8\ pN\ ms/nm$, $\lambda_2 = 10\ pN\ ms/nm$, $k_{gs} = 0.75pN/nm$, $k_{sp} = 0.6pN/nm$, $D = 61nm$, $k_B T = 4\ pN\ nm$, $\Delta G = 10k_B T$.

As the system is driven out-of-equilibrium by the non-conservative force and the effective temperature, there is a positive dissipation rate. The total entropy production rate can be calculated from the forces and their conjugated currents:[69,99]

$$\text{EPR}_{tot} = -\langle \dot{Q}_1 \rangle\left(\frac{1}{T} - \frac{1}{T_{eff}}\right) + \frac{\langle \dot{W}_{act} \rangle}{T_{eff}} \quad (4)$$

where $\langle \dots \rangle$ represents the steady state average. The steady state rate of the dissipated heat to the reservoir at temperature $T$ is $\langle \dot{Q}_1 \rangle = \langle (\partial V/\partial X_1) \circ \dot{X}_1 \rangle$, with $\circ$ being the Stratonovich product, and $\langle \dot{W}_{act} \rangle = -\langle F_{act} \circ \dot{X}_2 \rangle$ is the rate of work done by the active force.

## 3 Coarse-graining, lower bound on the total entropy production rate, and the mean dwell-time asymmetry factor

We used two approaches for spatial coarse-graining to discretize the continuous variable space (the trajectories of the tip position of the hair bundle, $X_1$) into discrete states: (i) dividing the continuous variable space equally into $N$ ($N$ = 3, 4, 5, 6, 7) coarse-grained states with the ratios 1:1:1, 1:1:1:1, 1:1:1:1:1, 1:1:1:1:1:1, and 1:1:1:1:1:1:1, respectively. This type of equal coarse-graining is only possible for a smooth trajectory for a particular choice of the driving parameter values (Figure S1, Supplementary Information, e.g. $F_{max} = 70\ pN, S = 1$, and $F_{max} = 80\ pN, S = 1$), (ii) diving the continuous variable space into unequal division, where $N$ ($N$ = 3, 4, 5, 6, 7) coarse-grained states correspond to dividing the trajectory with the ratios 1:1:1, $1:\frac{1}{2}:\frac{1}{2}:1$, $1:\frac{1}{3}:\frac{1}{3}:\frac{1}{3}:1$, $1:\frac{1}{4}:\frac{1}{4}:\frac{1}{4}:\frac{1}{4}:1$, and $1:\frac{1}{5}:\frac{1}{5}:\frac{1}{5}:\frac{1}{5}:\frac{1}{5}:1$, respectively, as shown schematically in Figure S2 of the Supplementary Information. This type of coarse-graining is better suited to track the irregular oscillations of the tip of the hair cell bundle for driving parameter values $F_{max}$= 90 pN, $S$ = 1, and $F_{max}$= 80 pN, $S$ = 1.5 etc. (see Supplementary Information, Figure S1).

We have two layers of coarse-graining in this manuscript: (I) one of the dynamical variables describing the system is decimated (In our example, the tip position of the hair bundle is observed, but the positions of the molecular motor are hidden) (II) we further coarse-grained the observed variable into a few discrete states.

Our system is coarse-grained such that the topology of the coarse-grained system is linear, without any cycles. The probability for a transition between the neighbouring states is non-zero, but the transition probability from one boundary state to the other boundary state is zero, and vice versa. For example, in a 3 coarse-grained system ($N$= 3, 1:1:1 spatial division), the probabilities of jumping from macro-state 2 to state 3 or 1 are both non-zero, whereas given the system is in state 1, the probability of finding it in state 3 in the next jump is zero, and vice versa. The waiting time distribution of the dwell time at state 2 depends, however, on the state visited before, whether it was state 3 or state 1, rendering the process a second order Markov process. Thus, we consider states composed of the current state, $i$, and previous state, $j$, i.e., $[i,j]$ when applying the KLD estimator. Similarly, the approach can be generalized to higher order semi-Markov processes.

Estimating dissipation is non-trivial in the absence of the observable currents, or flows, but as dissipated systems exhibit broken time-reversal symmetry, time irreversibility can be exploited to infer the out of equilibrium nature of the underlying process from the time series.[62] Martínez et al. developed an estimator based on the waiting time distributions containing information about irreversibility in hidden states even at the absence of visible transitions among the observed states. They applied the technique [62] for a partially hidden network where a subset of states are hidden, and a molecular motor system where the internal states are unresolved. In both cases, their estimator is able to predict a non-zero bound on the entropy production rate at the stalling driving force (the driving parameter value at which the current between the observed states vanishes).

To estimate the lower bound of the irreversibility, we use the KLD estimator[99,100], which relies on the broken time-reversal symmetry of the underlying waiting-time distributions.[62] Due to the presence of coupled hidden degrees of freedom, the jump process in the observed variable space becomes a second-order[62] semi-Markov. The jump probability depends on the previous state, the time since the last jump, and the final state. The last two conditions make the system direction-time dependent,[85] which means that the joint distribution of times and transitions ($\psi_{nn'}(t)$) cannot be written as a product of the probability distribution for a transition ($\Phi_{nn'}$) and the probability distribution of the time $t$ the system waits at the initial state $n$ ($\psi_n(t)$). As proved earlier,[62] the KLD estimator of the EPR for a semi-Markov process consists of two contributions: the affinity EPR (EPR$_{aff}$) and the waiting-time-distribution EPR (EPR$_{WTD}$). EPR$_{aff}$ accounts for the net current and the thermodynamic force of the system. It is sometimes called the "equivalent dissipation".[85] A non-Markovian system and its memoryless counterpart – a system with the same network topology generating Markovian sequence of states – have the same expression, but, the rate constants are replaced with the effective rate constants for the non-Markovian system. The affinity EPR is written as

$$\text{EPR}_{\text{aff}} = \frac{1}{\tau}\sum_{ijk} p(ijk) \ln \frac{p([ij]\to[jk])}{p([kj]\to[ji])} \quad (5)$$

where $p(ijk) = R_{[ij]}p([ij] \to [jk])$ is the probability to observe the sequence $i \to j \to k$. $R_{[ij]}$ denotes the normalized occupancy probability at the CG state $j$ given the previous CG state was $i$. The numerator and the denominator of the argument of the logarithmic function are of the form $p([ij] \to [jk])$, which denotes the probability that the system makes a transition from a CG state $j$ to a CG state $k$ given that the previous CG state was $i$. $\tau$ is the mean step duration given by $\tau = \sum_{ij} R_{[i,j]}\tau_{[i,j]}$, where $\tau_{[i,j]}$ is the mean time the system spends at a CG state $j$ given that the previous CG state was $i$. The sum is performed over all CG states ($i, j, k$). For the active hair bundle system, there is no contribution to the EPR from the affinity EPR, since the coarse-grained system is a linear chain of states.

The other component of the KLD estimator comes from the broken time-reversal symmetry in the waiting-time distributions and is obtained using the following equation:

$$\text{EPR}_{\text{WTD}} = \frac{1}{\tau}\sum_{ijk} p(ijk) D[\Psi(t|ijk)||\Psi(t|kji)] \quad (6)$$

Where $\Psi(t|ijk)$ denotes the probability density function of the time $t$ the system spends at a CG state $j$ before jumping to another CG state $k$, given that the previous CG state was $i$, i.e., for $i \to j \to k$ transition. The WTD estimator, EPR$_{\text{WTD}}$, or the "memory dissipation",[85] is the additional contribution that only exists for non-Markovian systems in contrast to their memoryless Markovian counterpart. It was shown that a semi-Markov process results in non-exponential waiting time distributions[101] which relate to memory.[85]

Since there is no net current in the observed variable space, the position of the hair-bundle tip, $X_1$, we use the KLD estimator [62] to calculate a lower bound on the total EPR. In order to apply this estimator, which was developed for discrete states, to a continuous variable system, we coarse-grain the observed variable into a few discrete states (a realization of 3 CG states in shown in Figure 1d), from which the lower bound is estimated by EPR$_{\text{WTD}}$, and study how the bound varies as a function of the number of coarse-grained states.

In order to demonstrate that a lower bound on the total EPR can be inferred from the WTD asymmetry in a system with second-order Markov process statistics with a linear topology having zero net current, we use a simple 6-state ($i = 1,2,3$ and $i' = 1', 2', 3'$, where states $i$ and $i'$ are indistinguishable) continuous time Markov chain (CTMC) model coarse-grained into a 3-state linear continuous time second-order semi-Markov system (observed states $1'', 2'', 3''$) as shown in Figure 2a. The net current in the 6-state model mimics the net current in the $X_1$-$X_2$ plane of the active hair bundle model Figure 1a, whereas the coarse-grained 3-state system resembles the coarse-grained, observed hair-bundle position, $X_1$. We simulated trajectories using the Gillespie algorithm[102] for $10^8$ steps, where after the decimation, we were left with approximately $10^6$ jumps. Figure 2b shows the difference in the distribution of the times the system waits at state $2''$ for an upward transition $(1'' \to 2'' \to 3'')$ and the corresponding downward transition $(3'' \to 2'' \to 1'')$. The non-exponential distribution originates from the non-Markovian statistics of the coarse-grained trajectory, whereas the difference between the distributions of the upward and downward waiting times originates from the nonequilibrium nature of the process.[62] Therefore, we can measure the irreversibility from the Kullback-Leibler divergence between the waiting time probability density functions EPR$_{\text{WTD}}$, for the coarse-grained system with zero EPR$_{\text{aff}}$ to provide a lower bound on the total EPR.

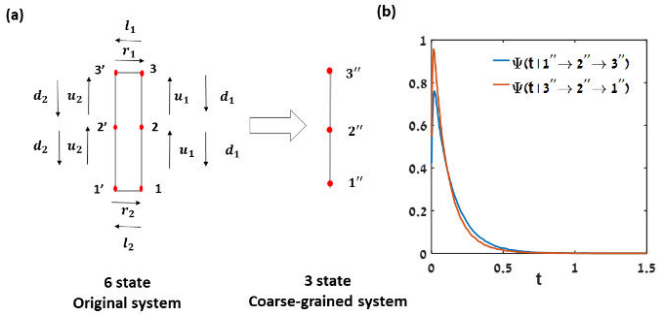

Figure 2 The $X_1 - X_2$ trajectory of the hair bundle system coarse-grained into a linear topology in $X_1$ state space after decimation of the $X_2$ states with zero net flux motivates to use KLD estimator of the waiting times: (a) The circles with the lines represents a 6 state system, which after decimation is reduced to a linear 3 state system, (b) non-zero contribution from the Kullback-Leibler divergence of the waiting time distributions: the distribution of the waiting times the system waits at CG state $2''$ for an ( $1'' \to 2'' \to 3''$) upward transition (blue solid line) and $(3'' \to 2'' \to 1'')$ the downward transition (red solid line) for the following parameter values: $u_1$ =10, $u_2$ =3, $d_1$ =2, $d_2$=4, $r_1$=3, $r_2$=3, $l_1$=1, $l_2$=1

For an example, the waiting time distributions for the hair bundle system at equilibrium ($F_{max} = 0\ pN, T = T_{eff}$) and at nonequilibrium condition driven according to Eq. 1 and Eq. 2 are shown in Figure 3a and Figure 3b, respectively. The distinguishability between the two WTD in the latter case b), results in a positive KLD which bounds the total EPR. The estimation of the $EPR_{WTD}$ improves with increasing the number of simulation steps (Figure 3c) as evident from the decreasing error and the plateauing of the estimation value for the active hair bundle model governed by Eq. 1 and Eq. 2 (c).[62]

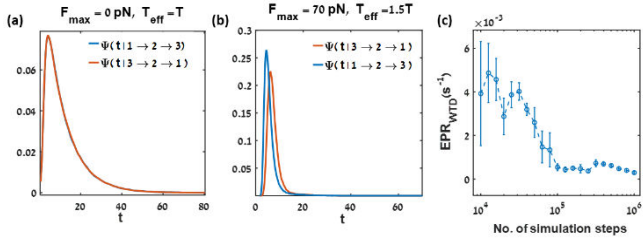

Figure 3 Entropy production rate estimation from the Kullback-Leibler divergence between the waiting time distributions for 3 equally spaced coarse-grained states of the active hair bundle's tip position: Probability density functions of times t that the system stays at state 2 for an upward transition (blue solid line), and for a downward transition (red solid line) for two different parameter values: (a) $F_{max}$=0 pN, $T_{eff} = T$, and $S = 1.5$, (b) $F_{max}$=70 pN, $T_{eff} = 1.5\ T$, and $S = 1.5$, (c) $EPR_{WTD}$ ($s^{-1}$) as a function of length of the simulation for $F_{max}$ =70 pN, $T_{eff}/T = 1.5$, and $S = 1.5$. The error bar at each point describes the standard error of the mean.

For a second-order semi-Markov process, the waiting time distributions are direction-time dependent. Thus, the mean dwell-times that the system spends at a particular state for the forward and the reverse transitions are not necessarily identical, and a deviation of their ratio from one provides information regarding the irreversible nature of the process.[78] We calculate the mean dwell-time asymmetry factor (MDAF), *i.e.*, the ratio between the means of the dwell time distributions ($\langle \tau_{k \to j \to i} \rangle$ or $\langle \tau_{kji} \rangle$) of times spent at a CG state $j$ before transitioning to $i$, given that it arrived from $k$, $k \to$ $j \to i,$ to the mean time the system spends at a CG state $j$ for a $i \to j \to k$ transition, ($\langle \tau_{i \to j \to k} \rangle$ or $\langle \tau_{ijk} \rangle$). The ratio between the mean times the system spends at a particular state before transitioning to another state and the mean times along the opposite direction ($\langle \tau_{kji} \rangle / \langle \tau_{ijk} \rangle$) being not equal to unity indicates a broken time-reversal symmetry in the system. To obtain the total MDAF for a system with $N$ coarse-grained states, we average over the individual MDAF of different transitions among the $N$ coarse-grained states. Therefore, the total MDAF equals $N^{-1} \sum \langle \tau_{kji} \rangle / \langle \tau_{ijk} \rangle$. The ratio stemming from the transitions among different coarse-grained states are plotted in the Supplementary Information (Figure S3).

In the following, we calculate the contribution of the $EPR_{WTD}$ from Eq. 6, and the effect of coarse-graining on the EPR and the MDAF, or the time-reversal symmetry breaking.

## 4 Effect of coarse-graining on the entropy production rate estimation and mean dwell-time asymmetry factor

We exploit the time-reversal symmetry breaking in the coarse-grained system to estimate the EPR. Since the affinity EPR vanishes, the signature of the irreversibility can only be tracked from the KLD between waiting time distributions, $EPR_{WTD}$.

First, The EPR estimate ($EPR_{WTD}$) values are calculated using Eq. 6 by coarse-graining the $X_1$ variable into $N$ CG states (where $N = 3,4,5,6,7$) by *equal* partitioning of the state space, and plotted as a function of $N$ (Figure 4a), for $F_{max} = 70\ pN$, $S = 1$, and $T_{eff}/T = 1.5$. The lower bound on the EPR estimate is improved with increasing resolution. The maximal value of $EPR_{WTD}/EPR_{tot} = 0.0013$ at 7 coarse-grained states. Moreover, the MDAF is plotted as a function of the number of the coarse-grained states (Figure 4b).

Next, we calculate the $EPR_{WTD}$ for several driving parameter values ($F_{max} = 70pN$, $F_{max} = 80\ pN$, $F_{max} = 90\ pN$, and $S = 0.5, 1, 1.5$) and for *unequal* spatial spacing of the coarse-grained states ($N = 3,4,5,6,7$). Both the estimate of the EPR (Figure 5a) and the mean dwell-time asymmetry factor (Figure 5b) increase with increasing spatial resolution. Indeed, the EPR estimate is correlated with the MDAF (Figure 5c), which is related to the non-Markovian nature of the process and the memory involved.[103]

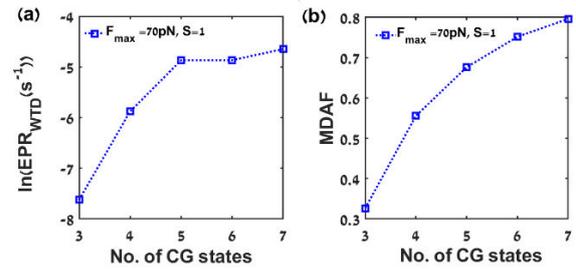

Figure 4 Entropy production rate estimation and the mean dwell-time asymmetry factor (MDAF) for equal spacing coarse-grining of $X_1$ trajectory: (a) $EPR_{WTD}$ ($s^{-1}$) (WTD estimate of the EPR) as a function of the number of CG states with equal spacing for parameter values $F_{max} = 70\ pN$, $S = 1$, and $T_{eff}/T = 1.5$ (b) MDAF (mean dwell-time asymmetry factor ) as a function of the number of CG states. The other parameter values are the same as mentioned in Figure 1. The lines are drawn to as a guide to the eye. The total EPR for this set of paramter values is 7.3312 $s^{-1}$.

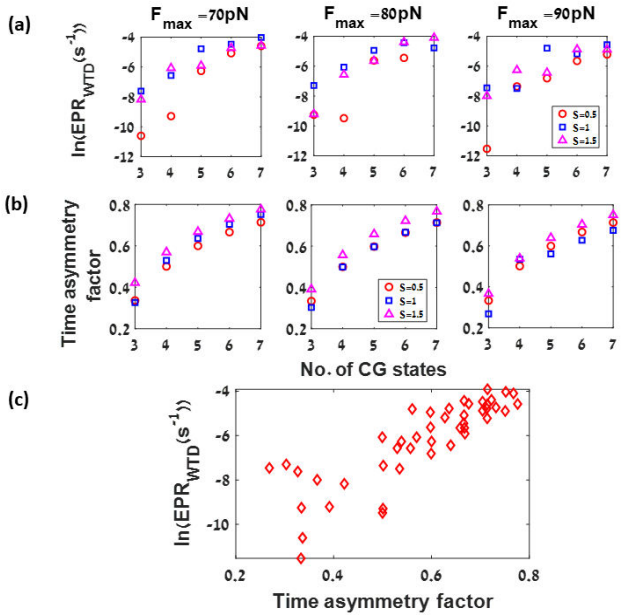

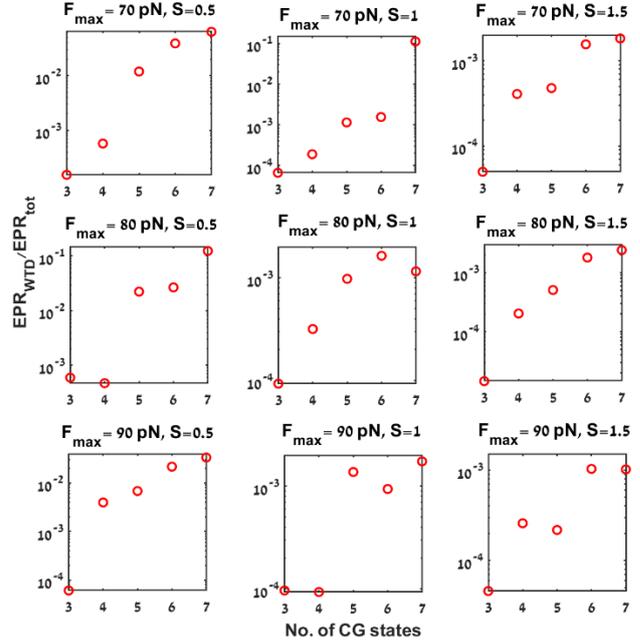

Figure 5 Effect of coarse-graining on the $EPR_{WTD}$ ($s^{-1}$) and the mean dwell-time asymmetry factor (MDAF): (a) $EPR_{WTD}$ ($s^{-1}$) as a function of number of CG states (3 CG: 1: 1: 1, 4 CG: 1: 1/2: 1/2: 1, 5 CG : 1: 1/3: 1/3: 1/3: 1, 6CG: 1: 1/4: 1/4: 1/4: 1/4: 1, and 7CG: 1: 1/5: 1/5: 1/5: 1/5: 1/5: 1) for different parameter values, (left) $F_{max}$=70 pN for S = 0.5,1,1.5, (middle) $F_{max} = 80$ pN for S = 0.5,1,1.5, and (right) $F_{max} = 90$ pN, S = 0.5,1,1.5.(b) The MDAF as a function of number of CG states for different parameter values: (left) $F_{max} = 70\ pN$, S = 0.5,1,1.5; (middle) $F_{max} = 80$ pN, S = 0.5,1,1.5; (right) $F_{max} = 90$ pN, S = 0.5,1,1.5. In both panels: red circle symbols correspond to $S = 0.5$, blue square symbols correspond to $S = 1$, and magenta triangle symbols correspond to $S = 1.5$. (c) The values of $EPR_{WTD}$ as a function of the MDAF for all transitions and all parameter values as mentioned earlier. The other parameter values used in these figures are mentioned in Figure 1.

As we mentioned $EPR_{WTD}$ was calculated for equal (Figure 4) and unequal (Figure 5) partitioning of the observed trajectory. For a certain driving parameter values at which the trajectories are not that smooth or regular. In that case, the equal partition of the trajectory space of the observed variable would lack enough statistics for the boundary states in the time series. Therefore, we consider unequal spatial partitioning of the trajectory.

To assess the tightness of the bound, we compare the ratio between $EPR_{WTD}$ estimates and the total EPR ($EPR_{tot}$) calculated for different driving parameter values, $F_{max} = 70\ pN, 80\ pN, 90\ pN$, $S = 0.5, 1, 1.5$, and for different coarse-graining levels (Figure 6), and find that the tightest bounds is obtained for 7 CG states ($N$ = 7), where the $EPR_{WTD}$ values are between 1 to 2 orders of magnitude smaller than the total EPR (Figure 6). The tightness of the bounds for unequal partitioning for 7 CG states are given in Table 1 of Supplementary Information.

## 5 Discussion

We study a nonequilibrium driven system with continuous variables following two coupled Langevin equations, where one of the degrees of freedom is observed and the other one is hidden.

We infer the irreversibility of the dynamics by coarse-graining the observed system variable into a few discrete states and applying the KLD estimator.[62] The coarse-grained *linear* system considered in our study is *non-Markovian*, but rather a second-order semi-Markov system, and the breaking of time-reversal symmetry is manifested in the difference between the non-exponential waiting time distributions of the forward and the reversed transitions among different coarse-grained states. The non-zero KLD measure between the forward and backward WTD reveals the underlying nonequilibrium dynamics.[62]

Figure 6 Tightness of the EPR bound ($EPR_{WTD}$) as a function of number of CG states: Ratio between the EPR estimates from the waiting time distribution ($EPR_{WTD}$ ($s^{-1}$)) and the total entropy production rate ($EPR_{tot}(s^{-1})$) for different parameter values. The coarse-graining corresponds to unequal divisions of the $X_1$ state space. The parameter values are $F_{max}$=70 pN, S = 0.5,1,1.5 (upper row), $F_{max}$=80 pN, S = 0.5,1,1.5 (middle row), $F_{max}$=90 pN, S = 0.5,1,1.5 (lower row). The other parameter values used in this figure are as mentioned in Figure 1.

Berezhkovskii *et al*.[103–106] discussed the case of low-resolution experimental observables in nonequilibrium systems, where the non-Markovian dynamics breaks time-reversal symmetry manifested in differences in the forward and backward waiting times. As suggested by several studies, [103–106] the time asymmetry in the active hair bundle system arises when the following two conditions hold: (i) reduced variable system follows non-Markovian statistics, and (ii) the system is out-of-equilibrium. Using a 6-state CTMC model which is coarse-grained into a linear 3-state system (Figure 2), (mimicking the hair cell bundle system with one degrees of freedom is decimated) we demonstrate that the resulting waiting time distributions calculated by the Gillespie algorithm[102] show characteristics of second-order semi-Markov statistics, and break time-reversal symmetry under nonequilibrium driving, and thus KLD estimator would be the good choice for the estimation of the EPR. The 6-state network decimated into 3 states mimics the coarse-graining of the $X_1$ trajectory into 3 coarse-grained states *(*Figure 1*d)*, in which a fundamental cycle is lost, and the contribution of the $EPR_{aff}$ vanishes. Indeed, we infer a lower bound on the total EPR, which can be calculated from the KLD between the distributions.

We calculate EPR estimates (EPR$_{WTD}$) of the continuous-space model system, an oscillating hair cell bundle, after coarse-graining the observed $X_1$ trajectory to equal (Figure 4a) and unequal (Figure 5a) spatial divisions. Comparing the results for a particular set of parameter values, $F_{max} = 70\ pN, S = 1$, and $T_{eff}/T = 1.5$, for which the trajectory is rather smooth and regular (see Supplementary Information, Figure S1). For the equal and unequal coarse-graining the lower bounds on the total EPR (*i.e.*, EPR$_{WTD}$/EPR$_{tot}$) are 0.0013, and 0.0024, respectively at parameter values $F_{max} = 70\ pN$, $S = 1$, and $T_{eff}/T = 1.5$.

The tightness of the lower bounds on the total EPR, *i.e.*, EPR$_{WTD}$/EPR$_{tot}$, is found to be 0.0013 for equal spatial division (Figure 4a) for $N = 7$ CG state at parameter value $F_{max} = 70\ pN$, $S = 1$, and $T_{eff}/T = 1.5$. Whereas, for unequal spatial division (Figure 6), EPR$_{WTD}$/EPR$_{tot}$ equals to 0.1244 for $N = 7$ coarse-grained states at $F_{max} = 80\ pN$, $S = 0.5$, $T_{eff}/T = 1.5$, respectively. The similar values of the EPR$_{WTD}$/EPR$_{tot}$ ratio results from the smooth nature of the $X_1$ trajectory at the chosen parameter set (as can be seen from Figure 1c) in contrast to the other parameter values (Supplementary Information, Figure S1). Equal spatial division for $N = 5, 6, 7$ coarse-grained states becomes challenging for parameter values that lead to very rugged trajectories due to the lack of statistics for the boundary states.

The inferred time-irreversibility and the EPR$_{WTD}$ estimate increase with finer spatial resolution, *i.e.*, larger number of CG states. Testing a wide range of parameter values, the EPR$_{WTD}$ lower bound is smaller by 1 to 2 orders of magnitude compared to the total ERP for the largest spatial resolution ($N = 7$) considered and unequal spacing of the observed $X_1$ trajectory, where the tightest bound, EPR$_{WTD}$/EPR$_{tot} \sim 0.1244$, is obtained for $F_{max} = 80\ pN$, $S = 0.5$, and $T_{eff}/T = 1.5$. All the ratios (EPR$_{WTD}$/EPR$_{tot}$) for 7 coarse-grained states are listed in Table 1 in the Supplementary Information.

## 6 Conclusions

In summary, the hair bundle system was used as a model to study the effect of coarse-graining on the lower bound on the total entropy production rate, the mean dwell-time asymmetry factor. The lower bound on the EPR was estimated using the underlying broken time reversal symmetry induced by the active force for a system with Langevin dynamics and zero net current along the reduced variable space. This approach can be applied to any system following Langevin dynamics with arbitrary number of observed and hidden states, to quantify the deviation from thermal equilibrium manifested in irreversibility of the observed degrees of freedom.

## Author Contributions

A.G. and G.B. designed research, performed research, analysed data, and wrote the paper.

## Conflicts of interest

There are no conflicts to declare.

## Acknowledgements

G. Bisker acknowledges the Zuckerman STEM Leadership Program, and the Tel Aviv University Center for AI and Data Science (TAD). A. Ghosal acknowledges the support of the Pikovsky Valazzi scholarship.

This work is supported by the Air Force Office of Scientific Research (AFOSR) under award number FA9550-20-1-0426, and by the Army Research Office (ARO) under Grant Number W911NF-21-1-0101. The views and conclusions contained in this document are those of the authors and should not be interpreted as representing the official policies, either expressed or implied, of the Army Research Office or the U.S. Government.

## Notes and references

# Supplementary Information

# Inferring Entropy Production Rate from partially observed Langevin dynamics under Coarse-Graining

Aishani Ghosal[a] and Gili Bisker[*,a,b,c,d]

[a] Department of Biomedical Engineering, Tel Aviv University, Tel Aviv 6997801, Israel

[b] Center for Physics and Chemistry of Living Systems, Tel-Aviv University, Tel Aviv 6997801, Israel

[c] Center for Nanoscience and Nanotechnology, Tel-Aviv University, Tel Aviv 6997801, Israel

[d] Center for Light-Matter Interaction, Tel-Aviv University, Tel Aviv 6997801, Israel

*Corresponding author: Gili Bisker
Email: bisker@tauex.tau.ac.il

**Supplementary Information Text**

**Simulated trajectories.** We simulated Eq. 1 and Eq. 2 for the trajectories of the tip position of the hair bundle ($X_1$) – the observed variable. The trajectories for different values of the driving parameters ($F_{max}$, $S$, and keeping $T_{max} = 1.5\,T$) are plotted in *Figure S1*. These trajectories are later used to calculate the EPR bounds ($\text{EPR}_{\text{WTD}}$) on the total EPR ($\text{EPR}_{\text{tot}}$) and the mean dwell-time asymmetry factor (MDAF) as described in the main manuscript.

**Coarse-graining method.** We coarse-grain the trajectories of the observed variable ($X_1$) into 3, 4, 5, 6, and 7 discrete states by spatially dividing the $X_1$ state space to segments with ratios 1:1:1, $1:\frac{1}{2}:\frac{1}{2}:1$, $1:\frac{1}{3}:\frac{1}{3}:\frac{1}{3}:1$, $1:\frac{1}{4}:\frac{1}{4}:\frac{1}{4}:\frac{1}{4}:1$, $1:\frac{1}{5}:\frac{1}{5}:\frac{1}{5}:\frac{1}{5}:\frac{1}{5}:1$ respectively (as shown in *Figure S2* for *N*=3, 4, 5, 6). The parameters used for the calculations in *Figure S2* are: $F_{max} = 70\,pN$, $T_{eff} = 1.5T$, and $S = 1$. This coarse-graining is used for the results presented in Figure 5 and Figure 6 in the main manuscript.

**Mean dwell-time asymmetry factor (MDAF).** We calculate the total mean dwell-time asymmetry factor using $N^{-1}\sum \langle \tau_{kji} \rangle / \langle \tau_{ijk} \rangle$, where $N$ is the total number of coarse-grained states. The values of $(\langle \tau_{kji} \rangle / \langle \tau_{ijk} \rangle)$ from transitions among different coarse-grained states are plotted in *Figure S3*.

**Tightness of the bounds for different parameters:** We calculate the tightness of the bounds for unequal coarse-graining (as shown in *Figure S2*) for 7 coarse-grained states (*N*=7) at different driving parameter values, as shown in Table 1.

**Figure S1**

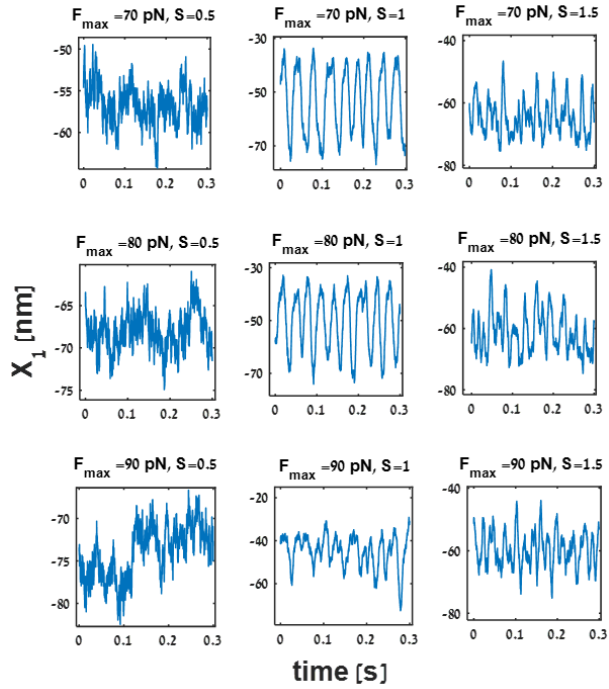

*Figure S1. The trajectories of the position of the tip of the hair bundle ($X_1$) as calculated by solving the coupled differential equations, Eq. 1 and Eq. 2 in the main text for different values of the parameter choices as a function of time. The driving parameter values are written in the subtitles with $T_{eff} = 1.5\,T$. All other parameter values are the same as mentioned in Figure 1 in the main text.*

**Figure S2.**

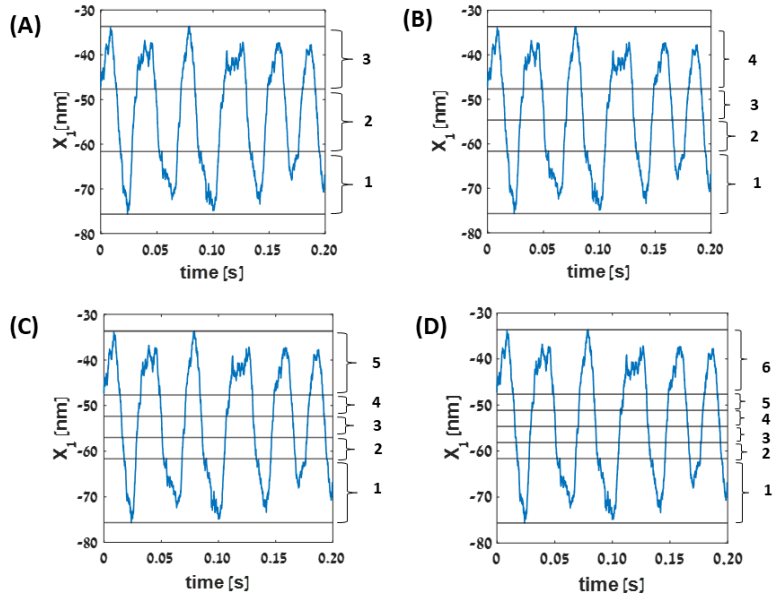

*Figure S2. Coarse-graining of the hair bundle tip position ($X_1$) for parameter values $F_{max} = 70\ pN$, $S = 1$, $T_{eff} = 1.5\ T$. The index numbers on the right side of each panel indicate the number of the coarse-grained states: (A) 3 CG states (1:1:1 division), (B) 4 CG states ($1:\frac{1}{2}:\frac{1}{2}:1$ division) (C) 5 CG states ($1:\frac{1}{3}:\frac{1}{3}:\frac{1}{3}:1$ division) and (D) 6 CG states ($1:\frac{1}{4}:\frac{1}{4}:\frac{1}{4}:\frac{1}{4}:1$ division). Panels (B), (C), and (D) show coarse-graining with unequal division of the $X_1$ trajectory as mentioned in the main manuscript. The top and bottom borders are determined by the extremum values of the $X_1$ parameter.*

**Figure S3.**

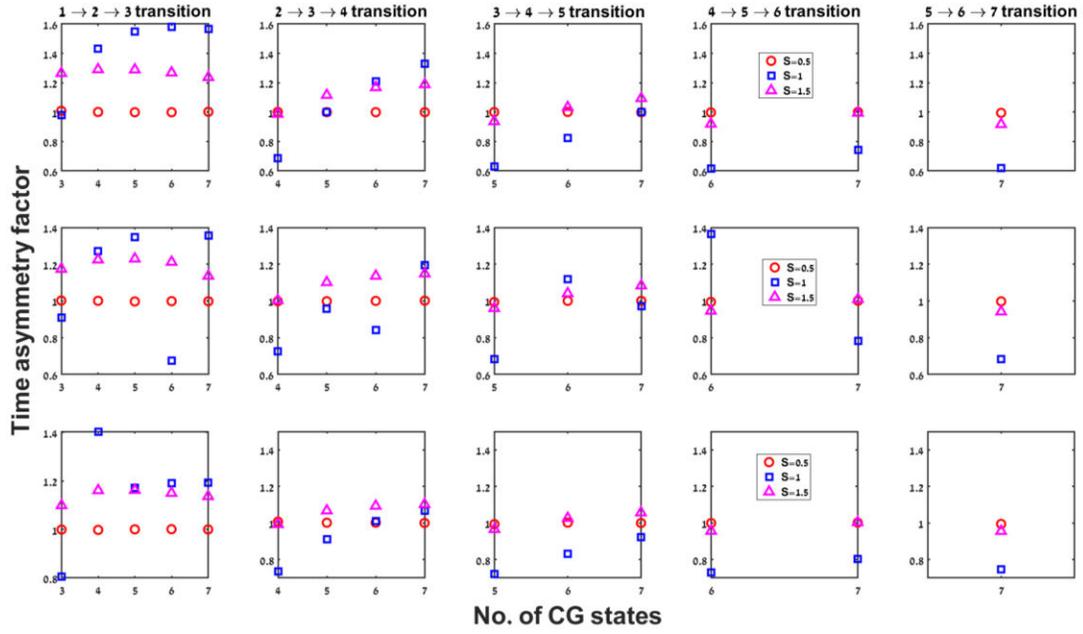

*Figure S3. The mean dwell-time asymmetry factors (MDAF, $\langle \tau_{kji} \rangle / \langle \tau_{ijk} \rangle$) as a function of the number of coarse-grained states for different transitions (shown in the subtitles) between the coarse-grained states for different parameter values: $F_{max} = 70\ pN$ (upper panel), $F_{max} = 80\ pN$ (middle panel), $F_{max} = 90\ pN$ (lower panel), $S = 0.5$ (red open circles), $S = 1$ (blue open square), $S = 1.5$ (magenta open triangle) ), and $T_{eff} = 1.5\ T$. All other parameter values are the same as mentioned in Figure 1 in the main text.*

**Table 1**

The tightness of the bounds for unequal coarse-graining for 7 coarse-grained states at different driving parameter values

| Driving parameter values | $EPR_{WTD}/EPR_{tot}$ |
|---|---|
| $F_{max} = 70\ pN, S = 0.5, T_{eff}/T = 1.5$ | 0.0666 |
| $F_{max} = 70\ pN, S = 1, T_{eff}/T = 1.5$ | 0.0024 |
| $F_{max} = 70\ pN, S = 1.5, T_{eff}/T = 1.5$ | 0.0018 |
| $F_{max} = 80\ pN, S = 0.5, T_{eff}/T = 1.5$ | 0.1244 |
| $F_{max} = 80\ pN, S = 1, T_{eff}/T = 1.5$ | 0.0012 |
| $F_{max} = 80\ pN, S = 1.5, T_{eff}/T = 1.5$ | 0.0024 |
| $F_{max} = 90\ pN, S = 0.5, T_{eff}/T = 1.5$ | 0.0335 |
| $F_{max} = 90\ pN, S = 1, T_{eff}/T = 1.5$ | 0.0017 |
| $F_{max} = 90\ pN, S = 1.5, T_{eff}/T = 1.5$ | 0.0010 |